\documentclass{aipproc}
\layoutstyle{6x9}

\usepackage{graphicx}
\usepackage{amsfonts}
\usepackage{amsmath}
\usepackage{verbatim}
\usepackage{color}

\newcommand{\half}{\frac{1}{2}}

\newcommand{\del}{\partial}
\DeclareMathOperator{\Tr}{Tr}

\newcommand{\eqn}[1]{ \begin{equation} #1 \end{equation} }
\newcommand{\field}[1]{\mathbb{#1}}

\newcommand{\Z}{\field{Z}}


\begin{document}%
\begin{flushright}
ADP-11-01/T723
\end{flushright}
\vspace{-20pt}

\title{A Novel Multiple-Time Scale Integrator for the Hybrid Monte Carlo Algorithm}
\author{Waseem Kamleh}{address={Special Research Centre for the Subatomic Structure of Matter and Department of Physics, University of Adelaide 5005, Australia.}}

\begin{abstract}
Hybrid Monte Carlo simulations that implement the fermion action using multiple terms are commonly used. By the nature of their formulation they involve multiple integration time scales in the evolution of the system through simulation time. These different scales are usually dealt with by the Sexton-Weingarten nested leapfrog integrator. In this scheme the choice of time scales is somewhat restricted as each time step must be an exact multiple of the next smallest scale in the sequence. A novel generalisation of the nested leapfrog integrator is introduced which allows for far greater flexibility in the choice of time scales, as each scale now must only be an exact multiple of the smallest step size.
\end{abstract}

\maketitle

\section{Introduction \label{sec:intro}}

Hybrid Monte Carlo(HMC)\cite{hmc} is the algorithm of choice for generating lattice gauge field configurations that include the effects of fermion loops. The main expense in such simulations is evaluating the contribution from the fermion determinant. A variety of improvements to the basic HMC algorithm have been developed in an effort to ameliorate this expense. Many of these variants involve the introduction of additional terms into the fictitious Hamiltonian. For example, Clark and Kennedy\cite{Clark:2006fx} introduce multiple pseudofermion fields into the Rational HMC algorithm, each contributing a fraction of the fermion determinant. Alternatively, Hasenbuch\cite{Hasenbusch:2001ne} uses the fermion matrix (with a heavier quark mass) as a preconditioner to the desired fermion action. This causes the fermion action to be split into two parts. Another example is the polynomial filtered HMC algorithm\cite{Kamleh:2005wg}, which uses short polynomial approximations to the inverse fermion matrix in order to separate the dynamics of the Hamiltonian. Multiple levels of filtering can be introduced, and this technique is also applicable to single flavour simulations, allowing the number of terms to be simulated in the Hamiltonian to grow to five or more. The Schwarz preconditioning technique employed by Luscher\cite{Luscher:2005rx} also makes use of a hierarchy of time scales.

The traditional way of dealing with a multiple time scale integration is to use the nested leapfrog algorithm by Sexton and Weingarten\cite{sexton-weingarten}. This scheme is quite restrictive however, as beginning with the finest time scale, the step-size for the next (coarser) time scale must be an exact multiple of the previous scale. This may prevent one from being able to choose the most efficient set of integration parameters, particularly if there are many time scales. Here we present an improvement of this scheme, which allows for greater freedom of choice, as each time scale must only be a multiple of the finest scale.

\subsection{Hybrid Monte Carlo Overview}

In this section we provide a brief overview of the Hybrid Monte Carlo algorithm\cite{hmc} in order to provide a framework for the introduction of our novel integrator. We wish to generate an ensemble $\{ U_i \}$ of representative gauge fields distributed according to the probability distribution $\rho(U_i) = e^{-S[U_i]},$ where the effective action for full QCD $S[U] = S_G[U] + S_F[U]$ can be divided into two parts, the gauge action $S_G[U]$ and the fermion action $S_F[U].$ Here we have assumed that the fermionic degrees of freedom have been integrated out in the usual way.

In the Hybrid Monte Carlo algorithm, the quantum lattice field theory is embedded in a higher-dimensional classical system through the introduction of a fictitious (simulation) time. The gauge field $U$ is associated with its (fictitious) conjugate momenta $P,$ and the classical system is described by the Hamiltonian,
\eqn{ H[U,P] = T[P] + S[U],}
where the fictitious kinetic energy is given by $T[P] = \sum_{x,\mu}\ \half \Tr P_\mu(x)^2.$
Given a configuration $U,$ a new gauge field $U'$ is generated by performing a HMC update $U \to U',$ which consists of two steps:
\begin{enumerate}
\item[(1)] \emph{Molecular Dynamics Trajectory:} Sample $P$ from a Gaussian ensemble. Integrate Hamilton's equations of motion to deterministically evolve $(U,P)$ along a phase space trajectory to $(U',P').$

\item[(2)] \emph{Metropolis step:} Accept or reject the new configuration $(U',P')$ with probability $\rho(U \to U') = \min(1,e^{-\Delta {H}}), \Delta {H} = {H}[U',P'] - {H}[U,P].$
\end{enumerate}

The discretised equations of motion are derived by requiring that the Hamiltonian be conserved along the phase space trajectory. We can express the equations of motion in terms of the time evolution operators induced by the kinetic and potential energy terms. The evolution operators that evolve the gauge field and its conjugate momenta forward a simulation time step $h$ are given respectively by
\begin{align}
V_T(h)&:\{U,P\}\to\{U\exp\big(i h P),P\}, \\
V_S(h)&:\{U,P\}\to\{U,P-hU\frac{\delta S}{\delta U}\}. 
\end{align}
Note that we demand that the evolution must preserve the $SU(3)$ property of the gauge field.

We must then combine these evolution operators into an overall evolution operator that is \emph{reversible. }The simplest such integration scheme is the leapfrog
\eqn{ V_H(h) = V_S(\frac{h}{2})V_T(h)V_S(\frac{h}{2}).}

After discretisation, for sufficiently small step sizes $h,$ the integration will conserve the Hamiltonian up to $O(h^2).$ 

\subsection{Multiple time-scales in molecular dynamics integrators}

If the action and thus the Hamiltonian is split into two parts $H_1$ and 
$H_2$,
\begin{equation}
  {H} = \underbrace{T[P] + S_1[U]}_{H_1} + 
             \underbrace{S_2[U]}_{H_2}
\end{equation}
then we define integrators for $H_1$ and $H_2$ as follows
\begin{equation}
\begin{matrix}
  V_{H_1}(h) = V_{S_1}(\frac{h}{2}) V_T(h) V_{S_1}(\frac{h}{2}), &
  V_{H_2}(h) = V_{S_2}(h).
\end{matrix}
\end{equation}
A compound integrator for the full Hamiltonian can be constructed by using a Sexton-Weingarten scheme\cite{sexton-weingarten}:
\eqn{
  V_H(h) = V_{H_2}(\frac{h}{2}) \left[ V_{H_1}(\frac{h}{m})\right]^m V_{H_2}(\frac{h}{2})
}
where $m \in \Z$. This nested leapfrog integrator effectively introduces two 
time-scales into the evolution, $h$ and $h/m$. Additional time scales may be introduced by repeating the nesting procedure.

\section{RESULTS}

We begin by assuming that our Hamiltonian consists of at least three terms,
\eqn{H = T + S_1 + S_2 + \ldots,}
where $T$ is the ``kinetic energy'' due to the conjugate momenta, and the terms $S_i$ implement the lattice QCD action. Typically we would choose $S_1 = S_G,$ that is $S_1$ is the gauge action. Then $S_2,S_3,\ldots$ will be the terms implementing the fermion action according to our algorithm of choice, be it mass-preconditioning or polynomial filtering and so on. The only thing we assume about the fermion action is that it is implemented  such that the only fields that are updated during the integration are the gauge field $U$ and its conjugate momenta $P$ (e.g. using pseudofermions).

For each scale we associate a timestep $h_i$ and a corresponding integer $N_i$ such that $h_i = 1/N_i.$ We require that $i=1$ corresponds to the scale at which the gauge field is updated. 

Assuming that $N_i > N_j$ for $i < j$ a nested leapfrog algorithm then requires that 
\eqn{ N_i | N_{i-1} \; \forall\ i > 1.\label{eq:divnl}} 
That is, $N_i$ must be a divisor of $N_{i-1},$ or equivalently, $h_i$ must be an exact multiple of $h_{i-1}.$ This means, for example, that each successive scale must be at least twice as coarse as the previous scale. It also may lead to being forced to choose a scale that is smaller than the one desired for a given time scale in order to simultaneously control the finite step-size errors as well as maintain the required arithmetic relation (\ref{eq:divnl}). The restriction of having to choose successive divisors for the various $N_i$ may not be the most efficient or flexible way of performing the molecular dynamics integration. 

\subsection{A generalised leap-frog integrator}

We present a generalised integration scheme in which it is only required that the integration scales satisfy the relation
\eqn{N_i | N_1 \;\forall\ i > 1.} 
This is of course equivalent to requiring that the step size $h_i$ is an exact multiple of $h_1.$

In a standard leapfrog algorithm, one alternates between updates $V_T$ to the gauge field $U$ and updates $V_S$ to the conjugate momenta. Let $V_i$ denote the update to $P$ corresponding to the action $S_i.$ Now, as the guide bosons are held fixed during an integration the updates $V_i$ only depend upon the gauge field. As the updates $V_i$ are additive to $P,$ it follows that the different $V_i$ commute:
\eqn{V_i(\frac{h_i}{2}) V_j(h_j) V_i(\frac{h_i}{2})  =  V_j(h_j) V_i(h_i) = V_i(h_i) V_j(h_j).}
Define the integers
\eqn{m_i = N_1 \div N_i}
to be the ratios of the scales. In order to construct our reversible integrator we first define a map
\begin{equation}
\Theta[V;m,k \in \mathbb{N}] = \left\{ 
\begin{aligned} &V\text{ if } m | k \\ 
&I\text{(the identity) otherwise.} 
\end{aligned}\right.
\end{equation}
Let $m_T$ be the lowest common multiple of $\{ m_i \},$ and let $h_T$ be the smallest time step (in our case $h_T = h_1).$ Then our integrator is
\eqn{
 V(h) = \prod_i V_i(\frac{h_i}{2}) \times 
\prod_{k=1}^{m_T-1} V_T(h_T) \Big\{ \prod_i \Theta[V_i(h_i);m_i,k]\Big\} V_T(h_T) \\ \times \prod_i V_i(\frac{h_i}{2}),
}
where $h = m_T h_T$ is the total timestep taken by $V.$ The above expression is straightforwardly implemented in code. We demonstrate this with a pseudocode implementation here. Denote by $\{a \equiv b \mod m\}$ the usual notion of congruence modulo $m.$ Then we can implement the generalised integrator as follows.\\

\noindent\textbf{Pseudo-code for the generalised integrator:}

\begin{itemize}
\item For each term in the action $S_i$ perform an initial half-step $V_i(\half h_i)$ updating $P.$
\item Loop over $j=1$ to $N-1$ 
\begin{itemize}
\item Apply $V_T(h)$ to update $U.$
\item If $\{0 \equiv j \mod m_i\}$ apply  $V_i(h_i)$ to update $P$
\end{itemize}
\item Apply $V_T(h)$ to update $U.$
\item For each term in the action $S_i$ perform a final half-step $V_i(\half h_i)$ updating $P.$\\
\end{itemize}

 The advantage of the generalised integrator is that it allows finer control over the different scales. An analysis of the finite-step size errors for the generalised integrator is provided in the next section.

\subsection{Integrator Error Analysis}

We perform an error analysis of our generalised integrator for a simple choice of step-sizes, following the procedure in \cite{sexton-weingarten}. Given a Hamiltonian $H$ we can write the evolution operator for our system as $V(h) = \exp{(h\hat{H})},$ with step-size $h.$ Here we have defined $\hat{H}$ as the linear operator on the vector space of functions $f$ on phase space $(p,q)$ defined by the Poisson bracket
\eqn{
\hat{H}f = -\{ H, f \}
= \sum_i \left( \frac{\del H}{\del p_i}\frac{\del f}{\del q_i} - \frac{\del H}{\del q_i}\frac{\del f}{\del p_i} \right).
}
If we write the Hamiltonian as 
\eqn{H=T+S_1+S_2+S_3+S_4+\ldots}
then for each term in the Hamiltonian we can correspondingly define a
linear operator using the Poisson bracket relation above. We denote the linear operator associated with a given term in the Hamiltonian by adding a ``hat'' to the appropriate symbol.

Proceeding with the error analysis, we make use of the Baker-Campbell-Hausdorff result,
\eqn{
e^{\lambda \hat{A}}e^{\lambda \hat{B}}e^{\lambda \hat{A}}=\exp\Big( \lambda(2\hat{A}+\hat{B}) + 
\frac{\lambda^3}{6}([[\hat{A},\hat{B}],\hat{A}]+[[\hat{A},\hat{B}],\hat{B}])+O(\lambda^4)\Big)
}
and apply this to the generalised leapfrog integrator in the simple case of $H=T+S_1+S_2,$ where the time scale for each term in $H$ corresponds to a number of integration steps $N_T=6,N_1=3$ and $N_2=2$ respectively. The integrator for this simplest non-trivial case can be written as
\eqn{ V_H(h) = e^{\frac{h}{4}\hat{S}_2}e^{\frac{h}{6}\hat{S}_1}e^{\frac{h}{3}\hat{T}}e^{\frac{h}{3}\hat{S}_1}e^{\frac{h}{6}\hat{T}}e^{\frac{h}{2}\hat{S}_2}e^{\frac{h}{6}\hat{T}}e^{\frac{h}{3}\hat{S}_1}e^{\frac{h}{3}\hat{T}} e^{\frac{h}{6}\hat{S}_1}e^{\frac{h}{4}\hat{S}_2}.}
Repeated application of our BCH result allows us to deduce that the above expression can be written as
\begin{multline}
V_H(h) = \exp\Big( h \hat{H} + h^3\big(\frac{1}{48}[[\hat{S}_2,\hat{T}],\hat{T}]+\frac{1}{96}[[\hat{S}_2,\hat{T}],\hat{S}_2]+ \\
\frac{1}{216}[[\hat{S}_1,\hat{T}],\hat{S}_1]+\frac{1}{108}[[\hat{S}_1,\hat{T}],\hat{T}]+\frac{1}{36}[[\hat{S}_1,\hat{T}],\hat{S}_2]\big)\Big)
\end{multline}
From this expression we can immediately see that the error in the generalised integrator relative to the leading term is $O(h^2),$ just as for the regular leapfrog.

If we examine the individual leapfrog integrators corresponding to
\eqn{\begin{matrix}H_1 = T + S_1,& H_2 = T + S_2\end{matrix}}
we obtain
\begin{align}
V_{H_1}(h) &= e^{\frac{h}{6}\hat{S}_1}e^{\frac{h}{3}\hat{T}}e^{\frac{h}{3}\hat{S}_1}e^{\frac{h}{3}\hat{T}}e^{\frac{h}{3}\hat{S}_1}e^{\frac{h}{3}\hat{T}}e^{\frac{h}{6}\hat{S}_1} \\
&=\exp\Big( h \hat{H}_1 + h^3(\frac{1}{108}[[\hat{S}_1,\hat{T}],\hat{T}]+ \frac{1}{216}[[\hat{S}_1,\hat{T}],\hat{S}_1])\Big),
\end{align}
and
\begin{align}
V_{H_2}(h)&=e^{\frac{h}{4}\hat{S}_2}e^{\frac{h}{2}\hat{T}}e^{\frac{h}{2}\hat{S}_2}e^{\frac{h}{2}\hat{T}}e^{\frac{h}{4}\hat{S}_2} \\
&=\exp\Big( h \hat{H}_2 + h^3(\frac{1}{48}[[\hat{S}_2,\hat{T}],\hat{T}]+ \frac{1}{96}[[\hat{S}_2,\hat{T}],\hat{S}_2])\Big)
\end{align}

Hence we see that the only difference between the individual integrators and our generalised integrator is the cross term $[[\hat{S}_1,\hat{T}],\hat{S}_2].$ The algorithm is identical to a standard nested leapfrog in the case where $N_i|N_{i-1}.$

\section{Conclusions}

We have introduced a novel integration scheme that generalises the nested leapfrog scheme by Sexton and Weingarten\cite{sexton-weingarten}. The new scheme has the advantage that each term in the fictitious Hamiltonian may be assigned a time step that only need be an exact multiple of the finest time step. This is an improvement on the nested leapfrog, where each scale must be an exact multiple of the next smallest scale in the hierarchy.

Each term in the fictitious Hamiltonian will have a corresponding ``force term'' associated with it. The typical size of this force term leads one to choose an appropriate integration time scale for that term. The large the force term, the smaller the time scale that is required to keep the finite step-size errors under control. With the novel integration scheme introduced here one has more freedom to choose a time-scale that is appropriate to the force associated with a given term - in the nested leapfrog one may have been forced to choose a time scale that was smaller than needed due to the restrictions imposed by the size of the next smallest scale in the hierarchy.

In a HMC simulation with two degenerate flavours where the fermion action has been split into two (or more) pieces, we already have at least three different time scales: one for the gauge action, and two (or more) for the fermion action. With the development of multiple time scale algorithms such as polynomial filtered HMC\cite{Kamleh:2005wg} that are applicable to both two flavour and single flavour simulations, the number of terms in the Hamiltonian can grow quite large. It is in these cases that the novel scheme proposed here will become particularly useful.

\subsection*{Acknowledgments}

\noindent
The author would like to thank M. Peardon for discussions relating to the error analysis.



\end{document}